# New generalized unit distributions based on order statistics

Iman M. Attia *

Imanattiathesis1972@gmail.com ,imanattia1972@gmail.com

*Department of Mathematical Statistics, Faculty of Graduate Studies for Statistical Research, Cairo University, Egypt*

## Abstract

In the present paper, the author discusses derivation of unit distributions and derivation of the generalized form using the order statistics. The author discusses the Kumaraswamy as the smallest order statistics of the unit power distribution derived from the inverse Weibull distribution. The author discusses the unit Rayleigh distribution and how it can be generalized using the smallest, largest, and kth order statistics. Using the order statistics to generalize a distribution differs from other techniques like the power transformation and T-X family (transformed-transformer) method. For the discussed distribution, the author demonstrates the basic functions and properties with real data analysis.

## Keywords

Kumaraswamy distribution, Median based unit Rayleigh distribution, Unit distributions, order statistics. T-X family.

## Introduction

Unit distributions model and fit the proportionate data and ratio. They have wide applications in diverse fields like economics, finance, biology, medicine, hydrology, engineering, and sociology. A better understanding of the distribution fitting the data helps apply these data in various statistical applications like regression analysis, survival analysis, and time series analysis. Unit distribution can be derived by the following transformations $y = \frac{1}{1+w}, y = \frac{w}{1+w}, y = e^{-w}, y = \frac{e^w}{1+e^w}$ . There are many continuous distributions that model these data. Some of these distributions are the unit Gamma Lindley distribution (Karakaya & Sağlam, 2025), the unit half logistic geometric distribution (Ramadan et al., 2022), the logit truncated exponential skew logistic distribution (Pang et al., 2021), the arc-secant hyperbolic Weibull distribution (Korkmaz et al., 2023), the Vasicek distribution (Mazucheli et al., 2022), the logit slash distribution (Korkmaz, 2020), and the median-based unit Rayleigh distribution and the references therein (I. M. Attia, 2025a).

The generalization of the unit distribution can be achieved through different mechanisms like power transformation to obtain the power Johnson B (Cancho et al., 2020), power Generalized Johnson SB (Gallardo et al., 2022), and power unit inverse Lindley distribution (Gemeay et al., 2024). The generalization can also be conducted using the T-X family method (transformed-transformer mechanism) like the transmuted power unit inverse Lindley distribution (Eldessouky et al., 2025), Kumaraswamy generalized family of distribution (Tahir et al., 2020), generalized distribution based on T-Topp –Leone family of distributions (Sudsila et al., 2022), and the generalized unit half logistic geometric distributions (Nasiru et al., 2023).

The author discusses in this paper a different method for adding new parameter to the unit distribution using the general formula for the order statistics. Then the author demonstrates this by examples of different new distributions.

The paper is arranged into the following section. Section 1 explores the derivations of the different unit distributions discussed in the paper. Section 2 elaborates on some of the basic functions and properties of these distributions. Section 3 discusses the Maximum likelihood Estimators. Section 4 demonstrates real data analysis with expanded discussion. Section 5 comprehends the conclusions. Section 6 suggests the future work.

**Section one**: Derivation of the some unit distribution and their generalization:

**Preposition 1**: Kumaraswamy distribution has the following PDF seen in equation (1)

$$\alpha \beta y^{\beta-1}(1-y^\beta)^{\alpha-1} \quad 0 < y < 1, \quad \alpha \,\&\, \beta > 0 \qquad (1)$$

**Proof**: This distribution can be derived from the inverse Weibull distribution (IW) that has the following PDF and CDF in equation (2). Applying the transformation in equation (3) and the Jacobian in equation (4)

$$f_X(x) = \alpha \beta x^{-(\alpha+1)} e^{-\beta x^{-\alpha}}, \quad F_X(x) = e^{-\beta x^{-\alpha}}, \quad x > 0, \alpha \,\&\, \beta > 0 \qquad (2)$$

$$\text{let } y = e^{-\left(\frac{1}{x}\right)^\alpha} \text{ where } x > 0 \quad \text{so} \quad x = (-\ln y)^{\frac{-1}{\alpha}} \qquad (3)$$

$$\frac{dx}{dy} = \frac{-1}{\alpha}(-\ln y)^{-\frac{1}{\alpha}-1}\left(\frac{-1}{y}\right), \quad 0 < y < 1 \qquad (4)$$

Substituting equation (3) & (4) into (2) gives the so called unit power distribution with its PDF & CDF seen in equation (5) & (6).

$$f_Y(y) = \alpha\beta(-\ln y)^{1+\frac{1}{\alpha}} e^{-\beta(-\ln y)}\left(\frac{1}{\alpha}\right)(-\ln y)^{-\frac{1}{\alpha}-1}\left(\frac{1}{y}\right) = \beta y^{\beta-1}, 0 < y < 1, \beta > 0 \qquad (5)$$

$$F_Y(y) = y^\beta \quad , \quad 0 < y < 1, \beta > 0 \tag{6}$$

The general formula for the ith order statistics in sample size n is defined in equation (7)

$$f_{i:n}(y) = \frac{n!}{(i-1)!\,(n-i)!} \{F_Y(y)\}^{i-1} \{1 - F_Y(y)\}^{n-i} f_Y(y), y > 0 \tag{7}$$

Using the smallest order statistics formula, in other words, i=1, and substituting the parent distribution of unit power distribution derived from the IW distribution as previously explained gives the well-known Kumaraswamy distribution. Let $n = \alpha$, so equation (8) is the PDF of the Kumaraswamy. Therefore, Kumaraswamy distribution can be considered as generalization of the unit power distribution derived from the transformation of the IW distribution into unit power distribution. Because the unit power distribution in equation (5) has only one parameter ($\beta$). Using the smallest order statistics formula and substituting the PDF and CDF for this unit power distribution yield Kumaraswamy.

$$f_{1:n}(y) = \frac{n!}{(1-1)!\,(n-1)!} \{F_Y(y)\}^{1-1} \{1 - F_Y(y)\}^{n-1} f_Y(y), \quad 0 < y < 1$$

$$f_{1:n}(y) = \frac{n(n-1)!}{(n-1)!} \{1 - y^\beta\}^{n-1} \beta y^{\beta-1}, \quad 0 < y < 1$$

$$f(y) = \alpha \beta y^{\beta-1} \{1 - y^\beta\}^{\alpha-1}, \quad 0 < y < 1, \alpha \& \beta > 0 \tag{8}$$

What if we use the smallest order statistics and substitute the PDF and CDF of the IW distribution as the parent distribution, then apply the transformation and Jacobian in equations (3-4), this yields the same result, the well-known Kumaraswamy distribution.

**Preposition 2**: Fatima 1 distribution has the following PDF seen in equation (9).

**Proof**: Using the largest order statistics formula, substituting the PDF & CDF of the unit power distribution seen in equation (5), substituting n=i, and let $n = \alpha$, the new generalized form of unit distribution is shown in equation (9). Figure 1 illustrates the PDF and the hazard rate function of the distribution.

$$f_{n:n}(y) = \frac{n!}{(n-1)!\,(n-n)!} \{F_Y(y)\}^{n-1} \{1 - F_Y(y)\}^{n-n} f_Y(y), 0 < y < 1$$

$$f_{n:n}(y) = n\{F_Y(y)\}^{n-1} f_Y(y), \quad 0 < y < 1$$

$$f(y) = n\{y^\beta\}^{n-1} \beta y^{\beta-1} = \alpha \beta y^{\beta-1} \{y^\beta\}^{\alpha-1}, \quad 0 < y < 1, \quad \alpha \& \beta > 0 \tag{9}$$

The same distribution is acquired if using the formula of the largest order statistics with the PDF & CDF of the IW distribution as the parent distribution then applying the transformation and the Jacobian in equation (3-4).

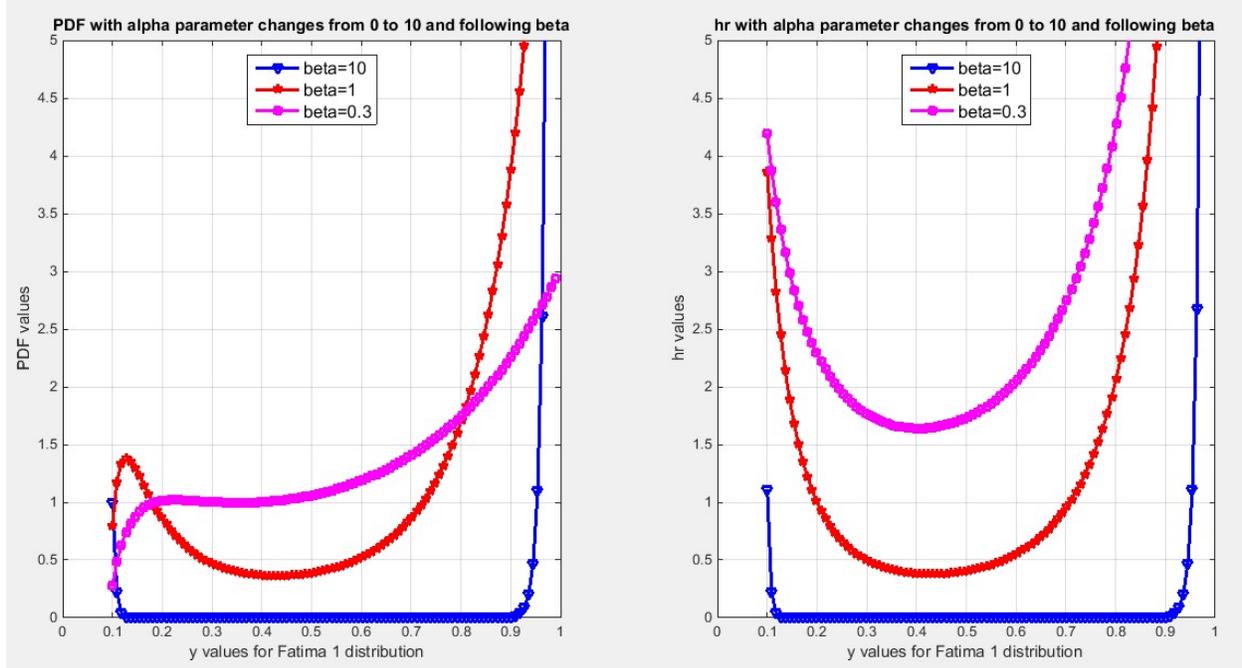

Fig.1 shows the PDF of Fatima 1 with J-shape (blue), increasing (pink) and the N shape (red) patterns. The hr function has J-shape to bathtub appearance. These figures are obtained with alpha values changing from 0 to 10 and beta has values of 10, 1, 0.3.

**Preposition 3**: Fatima2 distribution has the following PDF in equation (10).

**Proof**: Using the i$^{th}$ order statistics formula will result in generalized form of the unit power distribution with 3 parameters as shown in equation (10). Let $n = \alpha$. The same distribution is gained if using the i$^{th}$ order statistics, replacing the PDF & CDF of the IW as the parent distribution then applying the transformation and Jacobian of equations (3-4). Figure 2 shows the PDF and the hazard rate function.

$$f_{i:n}(y) = \frac{n!}{(i-1)!\,(n-i)!}\{F_Y(y)\}^{i-1}\{1 - F_Y(y)\}^{n-i} f_Y(y), \qquad 0 < y < 1$$

$$f_{i:n}(y) = \frac{n!}{(i-1)!\,(n-i)!}\{y^\beta\}^{i-1}\{1 - y^\beta\}^{n-i}\, \beta y^{\beta-1}, \qquad 0 < y < 1$$

$$f(y) = \frac{\Gamma n + 1}{\Gamma i!\, \Gamma(n-i+1)!}\{y^\beta\}^{i-1}\{1 - y^\beta\}^{n-i}\, \beta y^{\beta-1}, \qquad 0 < y < 1$$

$$f(y) = \frac{(\beta)\,\Gamma(\alpha+1)}{\Gamma(i)!\,\Gamma(\alpha-i+1)!}\, y^{\beta i - 1}\{1 - y^\beta\}^{\alpha - i}, \qquad 0 < y < 1 \qquad (10)$$

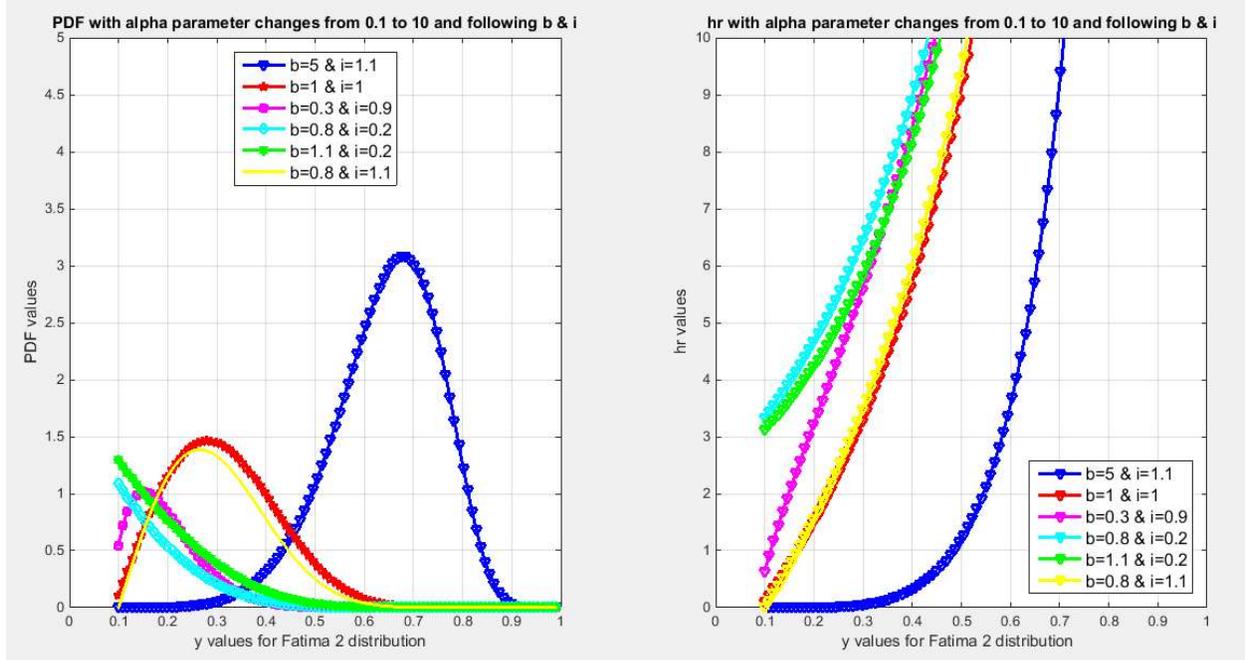

Fig. 2 shows the PDF of Fatima 2 distribution exhibiting right and left skewness and decreasing shape, while the hazard rate function depicts increasing pattern. The alpha values change from 0.1 to 10, and the i and beta values are as seen in the figure. The values of alpha should be larger than (i-1) to get valid CDF and hence valid hazard function.

Transforming the Rayleigh distribution which has PDF & CDF in equation (11) into unit Rayleigh distribution can be obtained by employing the transformation and the Jacobian seen in equation (12). The unit Rayleigh distribution has a PDF & CDF in equation (13) & (14) respectively.

$$F_W(w) = 1 - e^{\frac{-w^2}{\alpha^2}}, \quad f_W(w) = \frac{2w}{\alpha^2} e^{\frac{-w^2}{\alpha^2}}, \quad w > 0, \quad \alpha > 0 \quad (11)$$

$$y = e^{-w^2} \quad so \quad w = [-ln(y)]^{.5} \quad \frac{dw}{dy} = \frac{1}{2}[-ln(y)]^{-.5}\left(\frac{-1}{y}\right), \quad w > 0, \quad \alpha > 0 \quad (12)$$

$$f_Y(y) = \frac{1}{\alpha^2} y^{\left(\frac{1}{\alpha^2} - 1\right)}, \quad 0 < y < 1, \quad \alpha > 0 \quad (13)$$

$$F_Y(y) = y^{\frac{1}{\alpha^2}}, \quad 0 < y < 1, \quad \alpha > 0 \quad (14)$$

**Preposition 4**: Fatima 3 distribution had PDF seen in equation 15

**Proof**: Substituting equation (13) & (14) into the smallest order statistics formula gives the new generalized unit distribution in equation (15). Let $n = \beta$

$$f_{1:n}(y) = f_Y(y) = \frac{\beta}{\alpha^2} y^{(\frac{1}{\alpha^2}-1)} \left\{1 - y^{\frac{1}{\alpha^2}}\right\}^{\beta-1}, 0 < y < 1, \quad \alpha \,\&\, \beta > 0 \quad (15)$$

This result can also be obtained if we start with the largest order statistics formula and substitute Rayleigh distribution as the parent distribution with the PDF & CDF in equation (11) then applying the transformation and Jacobian in equation (12). Figure 3 depicts the PDF and the hazard rate function of the distribution.

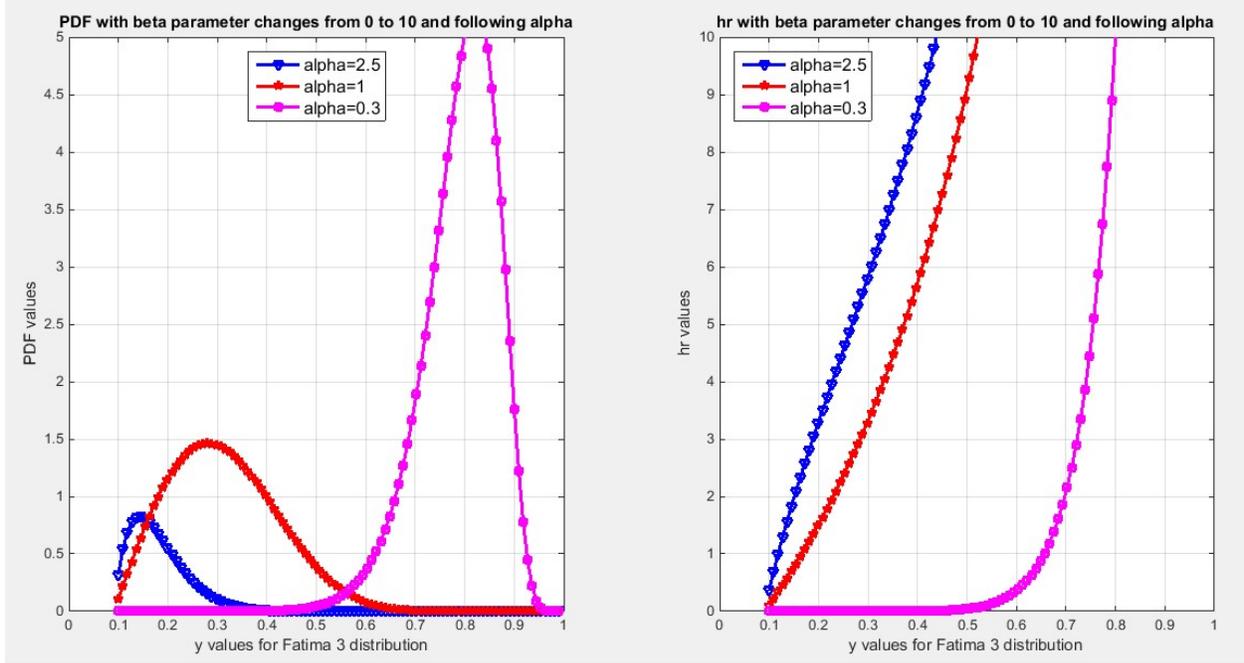

Fig. 3 shows the PDF of Fatima 3 distribution with right and left skewness according to the parameters values. The hr function is increasing. This figure is achieved with varying the beta values from 0 to 10 while the alpha parameter values are 2.5, 1, 0.3

**Preposition 5:** Fatima 4 distribution has the following PDF in equation 16

**Proof**: Substituting the PDF & CDF of equation (13-14) into the largest order statistics formula gives the new generalized unit distribution in equation (16). Let $n = \beta$.

$$f_{n:n}(y) = f(y) = \beta \left\{ y^{\frac{1}{\alpha^2}} \right\}^{\beta-1} \frac{1}{\alpha^2} y^{(\frac{1}{\alpha^2}-1)} = \frac{\beta}{\alpha^2} y^{(\frac{\beta}{\alpha^2}-1)}, 0 < y < 1, \; \alpha \,\&\, \beta > 0 \quad (16)$$

The same result is achieved if we start with the smallest order statistics formula and substitute the Rayleigh distribution with the PDF & CDF in equation (11) then utilizing the transformation and Jacobian in equation (12). Figure 4 illustrates the PDF and the hazard rate function of the distribution.

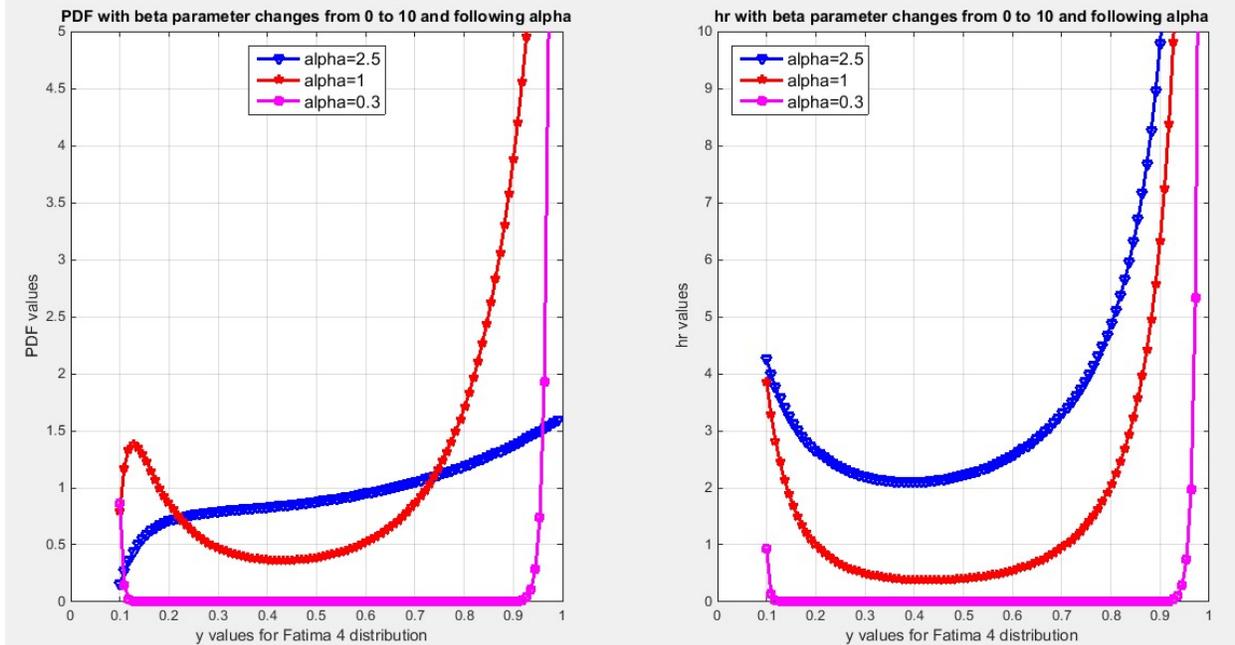

Fig. 4 shows the PDF of Fatima 4 with J- shape (pink), increasing (blue), and the N shape (red) patterns. The hr function has J appearance. These figures are obtained with beta changes from 0 to 10 and alpha has values of 2.5, 1, 0.3

**Preposition 6:** the Median Based Unit Rayleigh (Fatima 5) and its generalized form (Fatima 6 & Fatima 7) have the PDF in equations (17-19) respectively.

**Proof**: Using the PDF & CDF of Rayleigh distribution in equation (11) as the parent distribution , applying the transformation and Jacobian in equation (12) then substituting in the i$^{th}$ order statistics with i=2 and n=3 , gives the median based unit Rayleigh ( MBUR) distribution discussed by ( Attia,2025) as shown in equation (17). The author (I. M. Attia, 2025b) & (I. Attia, 2025) discussed the generalization of this distribution using the odd median order statistics as seen in equations (18-19).

$$f(y) = \frac{6}{\alpha^2}\left[1 - y^{\frac{1}{\alpha^2}}\right] y^{\left(\frac{2}{\alpha^2}-1\right)} , \ 0 < y < 1, \quad \alpha > 0 \quad (17)$$

$$f_y(y) = \frac{\Gamma(2n+2)}{\Gamma(n+1)\Gamma(n+1)} \frac{1}{\alpha^2}\left[1 - y^{\alpha^{-2}}\right]^n [y]^{\frac{n+1}{\alpha^2}-1}, n \geq 0, \alpha > 0, 0 < y < 1 \quad (18)$$

$$f_y(y) = \frac{\Gamma(n+1)}{\Gamma\left(\frac{n}{2}+\frac{1}{2}\right)\Gamma\left(\frac{n}{2}+\frac{1}{2}\right)} \frac{1}{\alpha^2}\left[1 - y^{\alpha^{-2}}\right]^{\frac{n-1}{2}} [y]^{\frac{n+1}{2\alpha^2}-1} , n \geq 1, \alpha > 0, 0 < y < 1 \quad (19)$$

**Section Two**: some basic properties

In this section, the author demonstrates some of the basic properties for each of the new unit distributions. This includes the raw moments and the quantile function.

**Preposition 7**: Fatima 1 has the following raw moments and quantile function seen in equations (20-21) respectively.

**Proof:**

$$E(y^r) = \int_0^1 \alpha\beta y^{\beta-1} y^r \{y^\beta\}^{\alpha-1} dy = \frac{\alpha\beta}{\alpha\beta + r} \qquad (20)$$

$$y = Q_Y(F_Y(y)) = F^{-1}(u) = u^{\frac{1}{\alpha\beta}} \quad , \quad 0 < u < 1, \qquad \alpha \,\&\, \beta > 0 \qquad (21)$$

**Preposition 8**: Fatima 2 has the following raw moments in equation (22) but it has no closed quantile function because it has no closed CDF.

**Proof:**

$$E(y^r) = \int_0^1 \frac{(\beta)\,\Gamma(\alpha + 1)}{\Gamma(i)\Gamma(\alpha - i + 1)} \, y^r \, y^{\beta i - 1} \{1 - y^\beta\}^{\alpha - i} \, dy$$

$$y^\beta = w \quad so \quad y = w^{\frac{1}{\beta}} \quad \& \quad \frac{dy}{dw} = \frac{1}{\beta} w^{\frac{1}{\beta} - 1} \,, \qquad 0 < w < 1$$

$$E(y^r) = \int_0^1 \frac{y^{\beta i - 1 + r} \,(\beta)\, \Gamma(\alpha + 1)}{\Gamma(i)\Gamma(\alpha - i + 1)} \{1 - y^\beta\}^{\alpha - i} dy = \frac{\Gamma(\alpha + 1)\,\Gamma\left(i + \frac{r}{\beta}\right)}{\Gamma(i)\,\Gamma\left(\frac{r}{\beta} + \alpha + 1\right)}, \quad 0 < y < 1, \quad \alpha, \beta \,\&\, i > 0 \qquad (22)$$

$$F_Y(y) = \frac{B_w(i, \alpha - i + 1)}{B(i, \alpha - i + 1)} \,, \quad i > 0 \,\&\, \alpha > i - 1$$

**Preposition 9**: Fatima 3 has the following raw moments and quantile function seen in equations (23-24) respectively.

**Proof:**

$$E(y^r) = \int_0^1 y^r \, \frac{\beta}{\alpha^2} \, y^{\left(\frac{1}{\alpha^2} - 1\right)} \left\{1 - y^{\frac{1}{\alpha^2}}\right\}^{\beta - 1} dy$$

$$y^{\frac{1}{\alpha^2}} = w \quad so \quad y = w^{\alpha^2} \quad \& \quad \frac{dy}{dw} = \alpha^2 w^{\alpha^2 - 1} \,, \qquad 0 < w < 1$$

$$E(y^r) = \int_0^1 y^r \, \frac{\beta}{\alpha^2} \, y^{\left(\frac{1}{\alpha^2} - 1\right)} \left\{1 - y^{\frac{1}{\alpha^2}}\right\}^{\beta - 1} dy = \frac{\Gamma(r\alpha^2 + 1)\,\Gamma(\beta + 1)}{\Gamma(r\alpha^2 + 1 + \beta)} \,, \qquad 0 < y < 1, \alpha \,\&\, \beta > 0 \qquad (23)$$

$$y = Q_Y(F_Y(y)) = F^{-1}(u) = \left(1 - [1-u]^{\frac{1}{\beta}}\right)^{\alpha^2}, \quad 0 < u < 1, \alpha \,\&\, \beta > 0 \quad (24)$$

**Preposition 10**: Fatima 4 has the following raw moments and quantile function seen in equations (25-26) respectively.

**Proof:**

$$E(y^r) = \int_0^1 y^r \frac{\beta}{\alpha^2} y^{\left(\frac{\beta}{\alpha^2}-1\right)} dy = \frac{\beta}{\beta + r\alpha^2}, \quad 0 < y < 1, \alpha \,\&\, \beta > 0 \quad (25)$$

$$y = Q_Y(F_Y(y)) = F^{-1}(u) = (u)^{\frac{\alpha^2}{\beta}}, \quad 0 < u < 1, \alpha \,\&\, \beta > 0 \quad (26)$$

## Section Three: Estimation Method (MLE)

**Preposition 11:** Fatima 1 has log of PDF in equation 27 and its partial derivative with respect to each parameter in equation (28) & (29)

$$l(\alpha, \beta | y) = n \log \alpha + n \log \beta + (\alpha\beta - 1) \sum_{i=1}^{n} \log y_i \quad (27)$$

$$\frac{\partial l}{\partial \alpha}(\alpha, \beta | y) = \frac{n}{\alpha} + \beta \sum_{i=1}^{n} \log y_i \quad (28)$$

$$\frac{\partial l}{\partial \beta}(\alpha, \beta | y) = \frac{n}{\beta} + \alpha \sum_{i=1}^{n} \log y_i \quad (29)$$

**Preposition 12:** Fatima 2 has log of PDF in equation 30 and its partial derivative with respect to each parameter in equation (31), (32) & (33)

$$l(\alpha, \beta, i | y) = n \log \beta + n \log \Gamma(\alpha + 1) - n \log \Gamma(i) - n \log \Gamma(\alpha - i + 1)$$

$$+ (\beta i - 1) \sum_{k=1}^{n} \log y_k + (\alpha - i) \sum_{k=1}^{n} \log\left(1 - y_k^\beta\right) \quad (30)$$

$$\frac{\partial l}{\partial \alpha}(\alpha, \beta, i | y) = n \psi(\alpha + 1) - n \psi(\alpha - i + 1) + \sum_{k=1}^{n} \log\left(1 - y_k^\beta\right) \quad (31)$$

$$\frac{\partial l}{\partial i}(\alpha, \beta, i | y) = -n \psi(i) + n \psi(\alpha - i + 1) + \beta \sum_{k=1}^{n} \log y_k - \sum_{k=1}^{n} \log\left(1 - y_k^\beta\right) \quad (32)$$

$$\frac{\partial l}{\partial \beta}(\alpha,\beta,i|y) = \frac{n}{\beta} + i\sum_{k=1}^{n}\log y_k - (\alpha - i)\sum_{k=1}^{n}\frac{y_k^\beta \log y_k}{1 - y_k^\beta} \tag{33}$$

**Preposition 13:** Fatima 3 has log of PDF in equation 34 and its partial derivative with respect to each parameter in equation (35) & (36).

$$l(\alpha,\beta|y) = n\log\beta - n\log\alpha^2 + \left(\frac{1}{\alpha^2} - 1\right)\sum_{k=1}^{n}\log y_k + (\beta - 1)\sum_{k=1}^{n}\log\left(1 - y_k^{\frac{1}{\alpha^2}}\right) \tag{34}$$

$$\frac{\partial l}{\partial \alpha}(\alpha,\beta|y) = \frac{-2n}{\alpha} - \frac{2}{\alpha^3}\sum_{k=1}^{n}\log y_k + \frac{2(\beta - 1)}{\alpha^3}\sum_{k=1}^{n}\frac{y_k^{\frac{1}{\alpha^2}} \log y_k}{1 - y_k^{\frac{1}{\alpha^2}}} \tag{35}$$

$$\frac{\partial l}{\partial \beta}(\alpha,\beta|y) = \frac{n}{\beta} + \sum_{k=1}^{n}\log\left(1 - y_k^{\frac{1}{\alpha^2}}\right) \tag{36}$$

**Preposition 14:** Fatima 4 distribution has log of PDF in equation 37 and its partial derivative with respect to each parameter in equation (38) & (39).

$$l(\alpha,\beta|y) = n\log\beta - n\log\alpha^2 + \left(\frac{\beta}{\alpha^2} - 1\right)\sum_{k=1}^{n}\log y_k \tag{37}$$

$$\frac{\partial l}{\partial \alpha}(\alpha,\beta|y) = \frac{-2n}{\alpha} - \frac{2\beta}{\alpha^3}\sum_{k=1}^{n}\log y_k \tag{38}$$

$$\frac{\partial l}{\partial \beta}(\alpha,\beta|y) = \frac{n}{\beta} + \frac{1}{\alpha^2}\sum_{k=1}^{n}\log y_k \tag{39}$$

## Section Four: Real Data Analysis and Discussion

The data sets are derived from the OECD, or Organization for Economic Co-operation and Development. https://stats.oecd.org/index.aspx?DataSetCode=BLI . It provides information on the economy, social events, education, health, labor, and the environment in the member countries. Matlab 2014 R was used for analysis where the MLE function utilizes the derivative free Nelder-Mead algorithm for optimization. The author analyzes the water quality indicator. This quality is measured through self-reported satisfaction or by tracking the availability of clean drinking water and the extent of water pollution. In this database, the water quality is expressed as percentage of population satisfied with the quality of water or subjective perception of water safety and cleanliness. It is also presented as the proportion of population with access to water that is free of contamination and available when needed. The dataset is 0.92, 0.92, 0.79, 0.90, 0.62, 0.82, 0.87, 0.89, 0.93, 0.86, 0.97, 0.78, 0.91, 0.67, 0.81, 0.97, 0.80, 0.77, 0.77, 0.87, 0.82, 0.83, 0.83, 0.85, 0.75, 0.91, 0.85, 0.98, 0.82, 0.89, 0.81, 0.93, 0.76, 0.97, 0.96, 0.62, 0.82, 0.88, 0.7, 0.62, and 0.72. Table 1 demonstrates the descriptive analysis of the water quality dataset. Figure (5) depicts the boxplot of the data. The data shows left skewness and mild platykurtic appearance.

Table 1: descriptive statistics of the water quality dataset

| min | Mean | Standard deviation | skewness | kurtosis | 25th quantile | median | 75th quantile | max |
|---|---|---|---|---|---|---|---|---|
| 0.062 | 0.8332 | 0.0972 | -0.6059 | 2.9144 | 0.7775 | 0.8300 | 0.9100 | 0.98 |

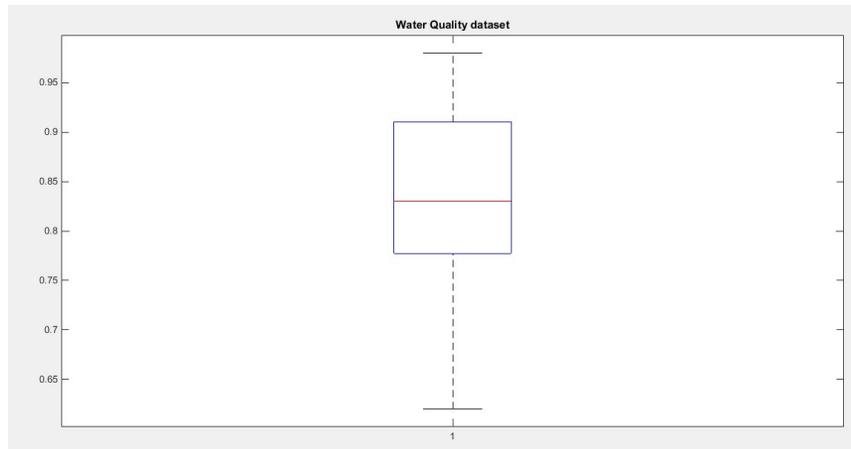

Fig. 5 shows the boxplot of water quality dataset.

Table 2: shows the beta distribution and the unit distributions derived from the IW distribution:

|  | Beta | | Kumaraswamy | | Fatima 1 | | Fatima 2 | | |
|---|---|---|---|---|---|---|---|---|---|
| $\alpha$ | 10.8716 | | 8.4271 | | 2.3004 | | 6.7786 | | |
| $\beta$ | 2.1667 | | 2.2817 | | 2.2928 | | 1.8642 | | |
| $i$ | - | | - | | - | | 5.6101 | | |
|  | 8.1698 | 1.2164 | 1.9713 | 0.5680 | 2.7745e+6 | -2.7653e+6 | Nan | Nan | Nan |
|  | 1.2164 | 0.2274 | 0.5680 | 0.2906 | -2.7653e+6 | 2.7561e+6 | Nan | Nan | Nan |
|  |  |  |  |  |  |  | Nan | Nan | Nan |
| LL | 40.6698 | | 40.4467 | | 34.9508 | | 40.6603 | | |
| AIC | -77.3397 | | -76.8933 | | -65.9015 | | -77.3205 | | |
| CAIC | -77.0239 | | -76.5775 | | -65.5857 | | -77.0047 | | |
| BIC | -73.9125 | | -73.4662 | | -62.4744 | | -73.8934 | | |
| HQIC | -76.0917 | | -75.6453 | | -64.6535 | | -76.0725 | | |
| H$_0$ | Fail to reject | | Fail to reject | | Fail to reject | | Fail to reject | | |
| P-value of KS | 0.8387 | | 0.7741 | | 0.0742 | | 0.07789 | | |
| KS-test | 0.0929 | | 0.0996 | | 0.1961 | | 0.0991 | | |
| CVM-test | 0.0398 | | 0.0483 | | 0.4182 | | 0.041 | | |
| AD-test | 0.3114 | | 0.3499 | | 2.2765 | | 0.3153 | | |

Table 2 demonstrates that all the competitors' distributions fit the data well as the null hypothesis test fails to reject the assumption that any of the distribution can generate the data. The P-value of the KS-test is significant for all of the fitted distributions. The beta distribution is the best to fit the data because it has the most negative AIC, CAIC, BIC and, HQIC followed by Fatima 2 distribution, then Kumaraswamy distribution and lastly Fatima 1 distribution. The variance of the estimated parameter obtained from fitting Fatima 1 and Fatima 2 is very large which may indicate correlation between the parameters and this requires more sophisticated methods to solve for parameter estimation.

Table 3: shows the unit distributions derived from the unit Rayleigh distribution:

| | Fatima 3 | | Fatima 4 | | Fatima 5 (BMUR) | Fatima 6 | | Fatima 7 | |
|---|---|---|---|---|---|---|---|---|---|
| $\alpha$ | 0.3445 | | 0.5199 | | 0.4776 | 0.4815 | | 0.4815 | |
| $\beta$ | 2.2817 | | 1.4257 | | - | 1.2081 | | 3.4162 | |
| | 0.0008 | -0.0116 | Nan | Nan | 0.0007217 | 0.000729 | 0.0037 | 0.000729 | 0.0073 |
| | -0.0116 | 0.2906 | Nan | Nan | | 0.0037 | 0.2140 | 0.0073 | 0.8558 |
| LL | 40.4467 | | 34.9508 | | 40.4976 | 40.6059 | | 40.6059 | |
| AIC | -76.8933 | | -65.9014 | | -78.9952 | -77.2118 | | -77.2118 | |
| CAIC | -76.5775 | | -65.5857 | | -78.8926 | -76.896 | | -76.896 | |
| BIC | -73.4662 | | -62.4744 | | -77.2816 | -73.7846 | | -73.7846 | |
| HQIC | -75.6453 | | -64.6535 | | -78.3712 | -75.9638 | | -75.9638 | |
| H$_0$ | Fail to reject | | Fail to reject | | Fail to reject | Fail to reject | | Fail to reject | |
| P-value of KS | 0.7741 | | 0.0742 | | 0.7789 | 0.822 | | 0.822 | |
| KS-test | 0.0996 | | 0.1961 | | 0.0991 | 0.0947 | | 0.0947 | |
| CVM-test | 0.0483 | | 0.4182 | | 0.0543 | 0.0416 | | 0.0416 | |
| AD-test | 0.3499 | | 2.2765 | | 0.3463 | 0.3214 | | 0.3214 | |

Table 3 shows that all the distributions fit the data well. Fatima 6 and Fatima 7 fit the data better than Fatima 5, followed by Fatima 3 and lastly Fatima 4. This is supported by the higher Log-likelihood and the most negative AIC, CAIC, BIC and HQIC. Variance –covariance of the estimated parameter gained from fitting Fatima 4 is very large which necessitate deploying other methods to manage this large variance. Figure (6-7) demonstrate the fitted PDF and CDF for the different unit distributions discussed in this paper.

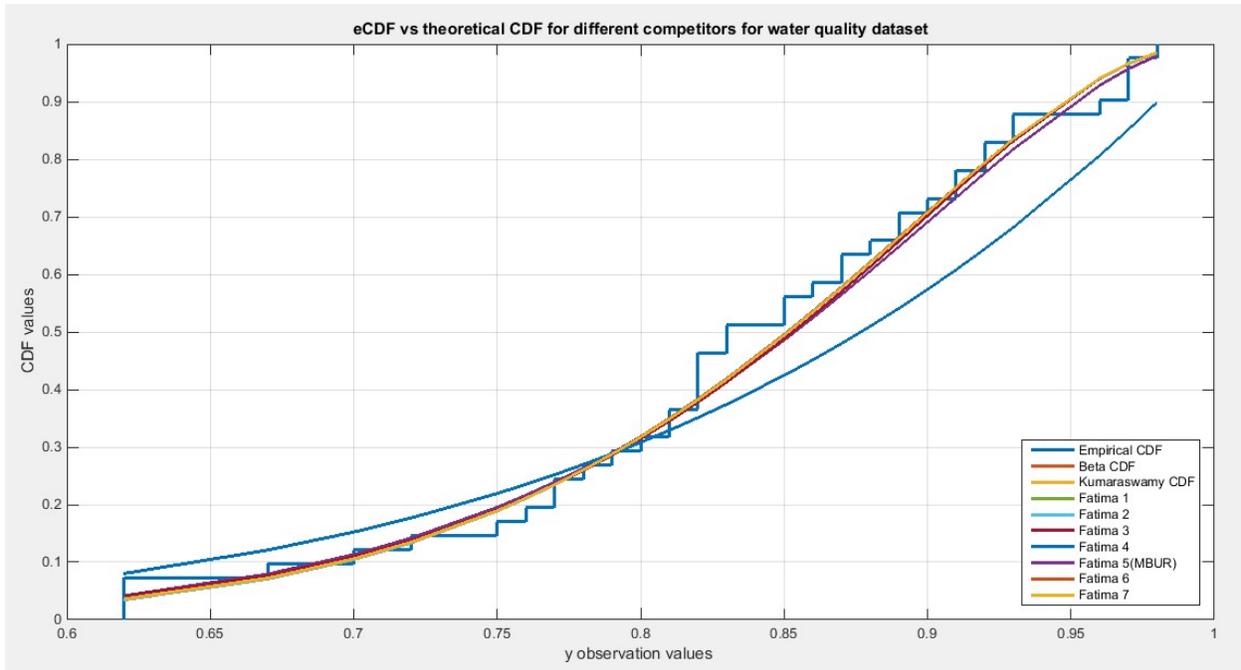

Fig. 6 shows the empirical CDF vs. theoretical CDF for the different unit distributions.

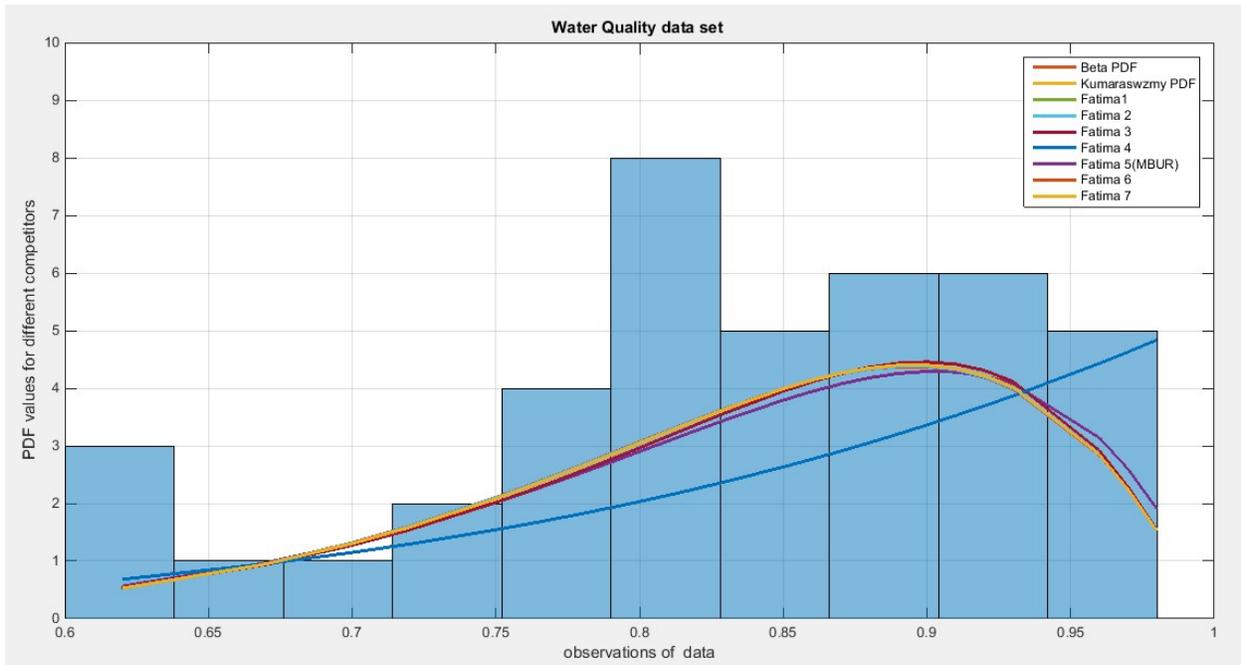

Fig. 7 shows the fitted PDF for the different distributions.

From the above discussion, all the distributions are more or less comparable to each other; they are nearly equal to fit the dataset except Fatima 1 and Fatima 4, as they fit the data with lesser value of Log-likelihood (34.9508). Although Fatima 1 is different from Fatima 4, however; the statistical indices like AIC, CAIC, BIC, HQIC, LL, KS, CVM, and AD are almost equal. The variance of the estimated parameters obtained from fitting Fatima 1, Fatima 2, and Fatima 4 distributions are very large. This may point to correlation between the parameters and so other methods are needed to mitigate this correlation for better parameter estimation and hence better construction of the confidence interval.

**Section five**: Conclusion

Generalization of the unit distribution is a challenge. The most frequent methods are the power transformation, methods utilizing Beta-generated or Kumaraswamy-generated families for a parent distribution of a random variable defined over the unit interval, or the transformer-transformed (T-X) family method. In this paper the author used a different method utilizing the general formula of the order statistics for a parent distribution of a variable defined over the unit interval. Kumaraswamy distribution is obtained from Inverse Weibull (IW) distribution by transforming this IW into unit power distribution and substituting the PDF and CDF of the new unit power distribution into the general formula of the smallest order statistics. We can also gain Kumaraswamy from general formula of the smallest order statistics substituting the IW PDF and CDF then applying the transformation of parent variable into a variable defined over unit interval as discussed in section 1. The largest order statistics formula and IW distribution can be used to derive Fatima 1. While the $i^{th}$ order statistics and the IW distribution are utilized to derive Fatima 2 distribution. By transforming the IW distribution into one-parameter unit power distribution then applying the general order statistics, a new shape parameter can be added to this unit distribution yielding different unit distributions according to the formula of the order statistics used.

Rayleigh distribution can be transformed into one-parameter unit Rayleigh then replacing the PDF and the CDF of this unit Rayleigh into the smallest order statistics yields Fatima 3 distributions. Moreover, Fatima 4 can be derived using the unit Rayleigh PDF and CDF as the parent distribution in the largest order statistics formula. Fatima 5 which is the Median Based Unit Rayleigh (MBUR) distribution originates from the general formula of the median order statistics with sample size equals 3 and generalization of this MBUR can be attained from using the general formula with an odd sample size to get Fatima 6 and Fatima 7. This method leads to an addition of a new parameter to the one-parameter unit Rayleigh distribution.

These new unit distributions give different estimators of their parameters with different variances. Some of these variances are very large pointing to the possibility of correlation between the newly added parameter and the existing old parameter. Hence, other methods should be deployed to mitigate this correlation and deflate the variance for better construction of confidence interval.

**Section Six**: future work

Estimation of the parameters of these unit distributions can be evaluated with other methods like least square method, weighted least square method, maximum product of spacing, Cramer Von Mises Estimator and Anderson Darling estimator. Bayesian inference can also be evaluated. Other properties of these distributions like entropy, Lorenz curve, Bonferroni curve, Gini index, probability weighted moments and mean residual life function are candidates for further study. Also stochastic ordering is a booster for further investigation. These distributions can be applied in analysis of censored data as well as survival data. They can be utilized in regression analysis as well as in time series analysis.

**Declarations:**
**Ethics approval and consent to participate**
Not applicable.
**Consent for publication**
Not applicable
**Availability of data and material**
Not applicable. Data sharing not applicable to this article as no datasets were generated or analyzed during the current study.
**Competing interests**
The author declares no competing interests of any type.
**Funding**
No funding resource. No funding roles in the design of the study and collection, analysis, and interpretation of data and in writing the manuscript are declared
**Authors' contribution**
AI carried the conceptualization by formulating the goals, aims of the research article, formal analysis by applying the statistical, mathematical and computational techniques to synthesize and analyze the hypothetical data, carried the methodology by creating the model, software programming and implementation, supervision, writing, drafting, editing, preparation, and creation of the presenting work.
**Acknowledgement**
Not applicable